\documentclass[aps,preprint]{revtex4}


\begin{document}

\preprint{$Revision: 1.00 $}

\title{W3 theory: robust computational thermochemistry in the kJ/mol accuracy
range}

\author{A. Daniel Boese, Mikhal Oren, Onur Atasoylu, Jan M. L. Martin{*}}

\email{comartin@wicc.weizmann.ac.il}

\homepage{http://theochem.weizmann.ac.il}

\affiliation{Department of Organic Chemistry, Weizmann Institute of Science, IL-76100
Re\d{h}ovot, Israel}

\author{Mih\'aly K\'allay and J\"urgen Gauss}

\affiliation{Institut f\"ur Physikalische Chemie, Universit\"at Mainz, D-55099 Mainz,
Germany}

\date{Received Oct. 31, 2003; Accepted Nov. 12, 2003 (306406JCP)}

\begin{abstract}
We are proposing a new computational thermochemistry protocol denoted W3 theory, as
a successor to W1 and W2 theory proposed earlier [Martin and De Oliveira, J. Chem. Phys. \textbf{111}, 1843 (1999)]. The new method is both more accurate overall (error statistics for total atomization energies
approximately cut in half) and more robust (particularly towards systems exhibiting significant
nondynamical correlation) than W2 theory. The cardinal improvement rests in an approximate 
account for post-CCSD(T) correlation effects.
Iterative $T_3$ (connected triple excitations)
effects exhibit a basis set convergence behavior similar to the $T_3$
contribution overall. They almost universally decrease molecular binding energies. Their inclusion in isolation yields less accurate results than 
CCSD(T) nearly across the board: it is only when $T_4$ (connected quadruple excitations)
effects are included that superior
performance is achieved. $T_4$ effects systematically increase molecular binding energies.
Their basis set convergence is quite rapid, and even CCSDTQ/cc-pVDZ scaled by an empirical factor
of 1.2532 will
yield a quite passable quadruples contribution. The effect of still higher-order excitations 
was gauged for a subset of molecules (notably the eight-valence electron systems): $T_5$
(connected quintuple excitations)
contributions reach 0.3 kcal/mol for the pathologically multireference $X~^1\Sigma^+_g$ state
of C$_2$ but are quite small for other systems.
A variety of avenues for achieving accuracy beyond that of W3 theory were explored, to 
no significant avail. W3 thus appears to represent a good compromise between accuracy and computational
cost for those seeking a robust method for computational thermochemistry in the kJ/mol
accuracy range on small systems.
\end{abstract}
\maketitle

\section{introduction}

Computational thermochemistry has come of age in recent years\cite{CioslowskiBook}.
The available techniques represent various trade-offs between accuracy
and computational cost. 

The {}``Gaussian-n'' (G$n$) family of methods\cite{Gn} 
first brought `black box' thermochemistry for small molecules in the
kcal/mol range: yet errors for individual systems can still exceed
the average over their training sets by as much as an order of magnitude.
G$n$ theory relies on relatively small basis sets, additivity approximations,
and empirical corrections. 

Similar remarks apply to the CBS (``complete basis set'') family
of methods by Petersson and coworkers\cite{CBS}, which involve intricate
combinations of pair correlation extrapolations and empirical corrections.

Some years ago, one of us proposed two new computational thermochemistry
protocols named W1 and W2 theory \cite{W1,W1book} that had the following
design goals:

\begin{itemize}
\item mean absolute error over various training sets in the kJ/mol range
\item worst-case errors in the 1 kcal/mol range, except for truly pathological
systems
\item completely devoid of empirical parameters
\item explicitly including all effects that affect molecular binding energies
in at least the kJ/mol range for first-and second-row systems, such
as core-valence correlation, scalar relativistic effects, and first-order
spin-orbit coupling
\item still be efficient enough for application to systems with up to six
heavy atoms on a fast commodity computer
\end{itemize}
An extensive validation study\cite{W1followup} revealed these goals
to be fundamentally met. Recently, an extension to systems with very
small valence-subvalence gaps (such as alkali and alkaline earth metal compounds) 
has been proposed\cite{Sullivan}. Yet Ref.\cite{W1followup}, 
and our general experience, revealed two main Achilles' heels to the method:
\begin{enumerate}
\item As the nonrelativistic parts of W1 and W2 theory both represent extrapolations\cite{l4,Halkier}
to the CCSD(T) basis set limit, the methods are intrinsically prone
to failure for systems suffering from moderate to strong nondynamical
correlation effects
\item The scalar relativistic treatment is based on one-electron Darwin
and mass-velocity corrections\cite{DMV}. While this approach is easily implemented
and expected to work well for first-and second-row systems, application
of W1 and W2 theory to heavier element systems will require
a more rigorous relativistic treatment such as the Douglas-Kroll-Hess\cite{DK,Hess}
approximation.
\end{enumerate}
In the present paper, we shall investigate these and some ancillary
issues, focusing particularly on CCSD(T) insufficiency. We shall propose
a new member of the W$n$ family called W3 theory, which should be
capable of handling cases where W1 and W2 theory fail.  Furthermore, we
will report on some avenues we explored in seeking further improvements
compared to W3 theory.

\section{Computational Details}

Electronic structure calculations at the CCSD (coupled cluster with all single and
double substitutions\cite{Pur82}) and CCSD(T) (CCSD with 
quasiperturbative triple excitations\cite{Rag89,Wat93}) levels
were carried out using MOLPRO 2002.6 \cite{MOLPRO} running on an
Intel/Linux cluster in our group. Electronic structure calculations
at the CCSDT (coupled cluster with all single, double, and triple substitutions), 
CCSDTQ (coupled cluster with all single, double, triple and quadruple
substitutions), CCSDTQ5 (ditto with added connected quintuple substitutions) 
and full configuration interaction (FCI) levels
were carried out using the generalized CI/CC code developed by
one of us\cite{kallay1,kallay2,kallay3}. The latter was interfaced
to the atomic orbital integrals, self-consistent field (SCF), and integral transformation parts
of the Austin/Mainz version of ACES II \cite{ACES} which was also
itself employed for some of the CCSDT calculations. ROHF (restricted open-shell
Hartree-Fock) reference determinants were used throughout for open-shell
systems: the definition of the ROHF-CCSD(T) energy according to Ref.\cite{Wat93}
was employed throughout. All calculations were carried out using the `frozen core
approximation', except those using core-valence correlation basis sets.

Most basis sets employed belong to the correlation consistent family of Dunning
and coworkers\cite{ccECC}. Unless indicated otherwise, we have combined the
regular cc-pVXZ (correlation consistent polarized valence X-tuple zeta\cite{Dun89}) basis
set on hydrogen with aug-cc-pVXZ ([diffuse function] augmented cc-pVXZ\cite{Kendall92}) on 
B--Ne and, on Al--Ar, the aug-cc-pV(X+$d$)Z basis sets (aug-cc-pVXZ with additional high-exponent
$d$ function) of Dunning, Peterson, and Wilson\cite{cc-pVn+dZ}. 
For convenience, we will denote this combination by the abbreviation AVXZ throughout
the present paper. The abbreviation PVXZ will refer to the combination of
regular cc-pVXZ basis sets on H and B--Ne with cc-pV(X+$d$)Z on Al--Ar.

Most core correlation calculations were carried out
with the MTsmall (Martin-Taylor small\cite{W1}) basis set, which is a completely
uncontracted cc-pVTZ basis set with 2d1f high-exponent functions added. Additional
core correlation calculations were performed using the
cc-pwCVXZ (correlation consistent polarized weighted core-valence X-tuple zeta)
basis sets of Peterson and Dunning\cite{pwCVnZ}. 

In a slight departure from W2 theory, and for consistency with the other
basis sets used, reference geometries were obtained at the CCSD(T)/cc-pV(Q+d)Z
level. Zero-point vibrational energies (ZPVEs), obtained from experimental or 
high-level ab initio harmonic frequencies and anharmonic corrections, were
taken from Ref.\cite{W1} unless indicated otherwise.

Unless indicated otherwise, extrapolations to the infinite basis set
limit for correlation energies are carried out using the same simple
formula\cite{Halkier} employed in W2 theory\cite{W1}, $E(L)=E_\infty+a/L^3$,
where $L$ is the maximum angular momentum represented in the basis set
(2 for AVDZ, 3 for AVTZ, 4 for AVQZ, 5 for AV5Z, and 6 for AV6Z). This 
formula is based on the leading term in the partial wave expansion of 
singlet-coupled pair energies\cite{Kut92}. For the
SCF energy, the same $E(L)=E_\infty+a/L^5$ as in W2 theory was employed.

\section{Initial results and discussion}

\subsection{Importance of connected quintuple and higher excitations}

Ruden et al.\cite{TrygveQuadruples} noted that connected quintuple
excitations, i.e. CCSDTQ5 - CCSDTQ, account for up to 0.3 kJ/mol to
the dissociation energy of N$_2$ in a cc-pVDZ basis set.
Bartlett and coworkers\cite{bartlettT5} noted that connected quintuples
contribute as much as 1 cm$^{-1}$ to the harmonic frequency of N$_2$. While explicit
inclusion of connected quintuples would be computationally prohibitive
for all but the very smallest systems, we should at least verify whether
and to what extent connected quintuple and higher excitations could
become an issue. We considered (a) the atomic electron affinities (EAs);
(b) the dissociation energies of the
eight-valence electron diatomics C$_{2}$ , BN, BeO and MgO,
along with the B$_2$ diatomic. 

The largest FCI/AVDZ - CCSDTQ5/AVDZ difference, 0.07 meV, is found
for EA(O); all others are an order of magnitude less, or zero by definition.
We can safely state that an error of 70 $\mu$eV is of no concern
to most thermochemical applications, and hence that connected sextuple
and higher excitations can be safely neglected.

For the atomic EAs, the largest CCSDTQ5/AVTZ - CCSDTQ/AVTZ differences
are found for oxygen (0.87 meV) and nitrogen (0.55 meV). Turning to
the eight-valence electron systems, by far the largest contribution
there (0.32 kcal/mol) is for the pathologically multireference $X{}^{1}\Sigma_{g}^{+}$
state of the C$_2$ molecule. For the $a^{3}\Pi_{u}$ state this drops
to 0.14 kcal/mol; for the closed-shell singlet states of BN, BeO,
and MgO, we obtain +0.16, -0.11, and -0.04 kcal/mol, respectively.
Finally, connected quintuples contribute +0.08 kcal/mol to the binding
energy of B$_2$ and +0.13 kcal/mol to that of the CN radical.
 
As the asymptotic CPU time scaling of a CCSDTQ5 calculation is 
proportional to $n^5N^7$ (with $n$ the number of electrons correlated
and $N$ the number of virtual orbitals), a quintuples correction will be unfeasible in
all but the very smallest systems. Given that the resulting error is
in the fractional kJ/mol range, we consider its neglect an acceptable
price to pay for extending the applicability range of W3.

\subsection{Importance of connected quadruple excitations}

The importance of connected quadruple excitations, CCSDTQ - CCSDT,
as a function of basis set is displayed in Tables \ref{tab:T4IP} 
and \ref{tab:T4TAE}. 
Ruden et al.\cite{TrygveQuadruples} previously noted their importance for 
a much smaller set of systems.

First of all, connected quadruples systematically increase binding
energies as well as ionization potentials (IPs) and electron affinities (EAs).

Secondly, contributions in systems with significant nondynamical correlation
effects can be quite nontrivial. At the extrapolated basis set limit, we find contributions of 
2.31 and 2.05 kcal/mol, respectively, in the closed-shell singlet states of C$_2$ and BN,
and 1.81 kcal/mol for MgO. With just a PVDZ basis set, we find 1.75 kcal/mol for N$_2$O, 
1.71 kcal/mol for NO$_2$, and 3.21 kcal/mol for O$_3$. Clearly, contributions of that magnitude
are ignored at one's peril.

Thirdly, while basis set convergence is quite rapid, it is not uniform.
Convergence in systems like C$_2$ is definitely much slower than in,
e.g., H$_2$O. The case of C$_2$ is somewhat special as the zero-order wave
function is nearly biconfigurational, and connected quadruples relative
to the HF-SCF determinant are effectively double excitations with respect
to the dominant doubly excited determinant. 

Considering the asymptotic $n^{4}N^{6}$ CPU time scaling of a CCSDTQ
calculation, it would be very desirable if it could be carried out
in just a PVDZ basis set, perhaps with the use of a scaling factor
determined from the PVDZ/(basis set limit) ratio in a training
set of systems. (We chose the set of all systems in Table \ref{tab:T4TAE} for which
we were able to do CCSDTQ calculations in at least a PVTZ basis set.)
This approach would seem to work at least tolerably
well for many systems, but will not be universally applicable. Not
only in cases with a low-lying doubly excited state like C$_2$ will there
be a problem, but it can readily be seen from Table \ref{tab:T4TAE} that
the $T_{4}$ contributions for H$_2$O and HF go through a minimum as
a function of the basis set. (This is the case for the atomic electron
affinities of O and F as well, as well as for the $T_{4}$ contributions
to the atomic correlation energies. We suspect the issue to be specific
to these small and very highly electronegative elements.)

One reason why correlation consistent basis sets have overwhelmingly
supplanted atomic natural orbital basis sets is the much shorter integral
evaluation times for the former\cite{basis}
and that they tend to perform comparably for most applications. However,
the fractional integral evaluation time of a CCSDTQ calculation is
so ridiculously small that it may make sense to use the best possible
basis set for a given contracted size. We considered the averaged
ANO basis sets of Roos and coworkers\cite{WMR}, and found that the smallest
ANO contraction that yields acceptable results is {[}4s3p1d{]} (ANO431 for short). 
On the one
hand, we
find the {[}4s3p1d{]}/(basis set limit) ratio for the $T_{4}$ contribution
to be much more consistent, and hence it is much more amenable to
scaling. On the other hand, even the four additional basis functions per
nonhydrogen atom (relative to PVDZ) already make the O$_3$ molecule
nearly intractable on our presently available computational hardware. 
(The CCSDTQ/ANO431 calculation required 436 million determinants, compared to
a `mere' 111 million for CCSDTQ/PVDZ.)

In an attempt to eliminate the very costly CCSDTQ calculations, we considered
various continued-fraction and Pad\'e type approximations proposed by
Goodson \cite{Goodson}. Like a previous study of Feller and coworkers\cite{Feller2003},
we find these approximations to behave too erratically for practical
use, and we have abandoned them.

\subsection{Importance of higher-order connected triple substitutions}

Higher-order $T_{3}$ contributions --- as measured by the CCSDT -
CCSD(T) difference --- are tabulated in Table \ref{tab:T4IP} for atomic ionization
potentials and electron affinities, and in Table \ref{tab:T3TAE} for molecular total atomization
energies of our training set.

With a few exceptions (e.g., B$_2$ and CH) the contributions at the basis
set limit systematically reduce molecular binding energies. Thus,
as previously suggested by Bak et al.\cite{Bak2000}, the surprisingly good performance
of CCSD(T) (and, indeed, of W2 theory) is largely due to partial error
compensation between neglect of $T_{4}$ and iterative $T_{3}$ effects. 

Basis set convergence is considerably slower than for $T_{4}$. In
particular, contributions generally have a positive sign with the
PVDZ basis set and change sign as the basis set is expanded. Considering
that the contribution is itself 1--2 orders of magnitude smaller than
the (T) contribution to molecular binding energies, we can probably
get away with $E_\infty+a/L^3$ extrapolation\cite{Halkier} 
from AVDZ and AVTZ basis sets, thus keeping
CPU times for the CCSDT step (asymptotically proportional to $n^{3}N^{5}$)
within acceptable boundaries.

\subsection{Improved scalar relativistic correction}

In order to achieve greater robustness for heavier element systems,
we replaced the scalar relativistic treatment of W1 and W2 theory 
--- first-order Darwin and mass-velocity (DMV) corrections\cite{DMV} taken as 
expectation values from an averaged coupled pair functional\cite{ACPF} (ACPF)
wave function with the `Martin-Taylor small' (MTsmall) basis set\cite{W1} ---
 by a more rigorous one.
Specifically, the scalar relativistic contribution is taken as the
difference between the second-order Douglas-Kroll-CCSD(T)/aug$'$-cc-pRVQZ (ARVQZ for short)
and nonrelativistic
CCSD(T)/aug$'$-cc-pVQZ energies, where cc-pRVXZ stands for newly developed
relativistic correlation consistent X-tuple zeta basis sets\cite{Mikhal}.
(The prefix `aug$'$' denotes a basis set augmented with diffuse functions on the
main group elements but not on hydrogen\cite{Del93}.)
 A comparison between
this approach and the original DMV-ACPF/MTsmall treatment can be found
in Table \ref{tab:Mikhal}.

The bottom line is that the ACPF Darwin and mass-velocity approach,
while generally effective for first-and second-row systems, can actually
cause noticeable errors even for SO$_2$, and cannot be blindly relied
upon for heavier elements. 

Also, as seen from Table \ref{tab:Mikhal}, the relativistic correction with
the VQZ type basis sets is basically indistinguishable from the basis set limit.

\subsection{Improved extrapolation to the infinite-basis valence correlation
limit}

Klopper\cite{KlopperExtrap} proposed separate extrapolations of singlet-coupled
(as $E^S_\infty+a_SL^{-3}$) and triplet-coupled (as $E^T_\infty+a_TL^{-5})$ pair correlation
energies, corresponding to the leading terms of the partial wave asymptotic
expansions for such pairs\cite{Kut92}. The term linear in $T_{1}$ in the CCSD energy equation
(which is nonzero for open-shell CCSD calculations using semicanonical orbitals, such 
as done by MOLPRO\cite{T1energy}) is then
simply taken as that in the largest available basis set. 
Some results can be found in Table \ref{tab:extrap}. 

When extrapolating from AVQZ and AV5Z basis sets, separate extrapolation
systematically produces lower basis set limits than joint extrapolation.
Differences are by and large in the 0.1 kcal/mol range, but reach 0.16--0.18 
kcal/mol for HOCl, N$_2$O, and Cl$_2$, 0.2 kcal/mol
for CO$_2$ and OCS, and 0.3 kcal/mol for SO$_2$. When extrapolating from 
AV(5+d)Z and AV(6+d)Z basis sets, these discrepancies are greatly reduced:
this reflects the triplet-coupled pair energies being largely converged, leaving the
singlet-coupled pair energies to dominate convergence behavior. 
Furthermore, differences between the \{AVQZ,AV5Z\} and \{AV5Z,AV6Z\} extrapolated limits
are appreciable (e.g., 0.3 kcal/mol for Cl$_2$) using joint extrapolation,
and much smaller using separate extrapolation --- clearly suggesting the latter
to have more desirable convergence properties. On the other hand, using AVTZ and
AVQZ basis sets, the separate extrapolation is clearly performing more poorly than
the empirically damped (exponent 3.22) 
joint extrapolation used in W1 theory\cite{W1}.

As the (T) contribution is both smaller to begin with than the CCSD correlation
energy and converges more rapidly with the basis set\cite{Hel97},
standard W2w theory
extrapolates it from AVTZ and AVQZ  basis sets. (In this
manner, the largest basis set calculation in W2w is just a CCSD calculation
and can be carried out using integral-direct algorithms where necessary\cite{dirCCSD}.)
We considered the effect of extrapolating the (T) contribution from larger
AVQZ and AV5Z basis sets (Table \ref{tab:extrap}), and found
it to be below 0.1 kcal/mol in all cases and below 0.05 kcal/mol in most species.

As to the SCF component, the effect of extrapolating from AV5Z and
AV6Z basis sets is negligible at our target accuracy level, with the
notable exception of SO$_2$ where inner polarization functions are known to
be very important\cite{so2}. We attempted SCF calculations in even larger
basis sets than aug-cc-pV(6+d)Z (particularly aug-cc-pV6Z+2d1f), and find out best Hartree-Fock
limit to be 121.93$\pm$0.04 kcal/mol, in between the \{AVQZ,AV5Z\} and \{AV5Z,AV6Z\}
extrapolated values.

Finally, we considered basis set superposition error (BSSE). 
Among
the different many-body generalizations\cite{Wel83,Opus3,AlmBSSE}
of the counterpoise correction\cite{Boy70}, we have followed the
'site-site function counterpoise' definition of Wells and Wilson\cite{Wel83}. 
The results are given in Table \ref{tab:BSSE}. We note that valence BSSEs
are fairly noticeable for the individual 
basis sets up to even the AV6Z level, 
but are largely annihilated by the extrapolation.

\subsection{Improved inner-shell correlation contribution}

In the original W1/W2 paper, it was established that connected triple
excitations are quite important (relatively speaking) in the core-valence
contribution to molecular binding energies. As a result, CPU times
in especially W1 calculations on second-row molecules and large first-row molecules
are dominated
by the inner-shell correlation step, and we had a vested interest
in keeping the core correlation basis set as small as possible. The
smallest basis set that could reliably reproduce them was found to
be what we termed the MTsmall basis set\cite{W1}. 
As we are 'tightening the screws' everywhere else, it makes sense
to explore the importance of better core correlation basis sets, especially
considering the in any case steep computational cost of the CCSDTQ
valence calculations.

Core-valence correlation contributions with the core-valence weighted\cite{pwCVnZ}
aug$'$-cc-pwCVTZ and aug$'$-cc-pwCVQZ
basis sets, as well as extrapolations to the infinite-basis limit,
can be found in Table \ref{tab:CORE}. In addition, we considered the effect
of basis set superposition error on the inner shell contribution,
following a suggestion by Bauschlicher and Ricca\cite{BauRicSO2}
that it might become quite important for second-row systems. 

We found a serious issue with BSSE for SO$_2$ (0.85 kcal/mol with the
smaller basis set), but even here simple $a+b/L^{3}$ extrapolation
basically eliminates the problem. 

\subsection{Use of Wilson's second-row basis sets}

The original W1 and W2 methods added high-exponent $2d1f$ sets to second-row
basis sets in order to cope with polarization of the (3s,3p) inner loops\cite{BauPar95,Mar98,so2}.
These basis sets do guarantee saturation of the HF-SCF energy even in extreme
cases like SO$_3$ (where inner polarization contributes 10 kcal/mol to the
HF-SCF binding energy even with an aug-cc-pVQZ basis set\cite{so3tae}). Recently,
however, Wilson and coworkers\cite{cc-pVn+dZ} published new so-called cc-pV($n$+d)Z
basis sets that are designed to cope with the phenomenon in a consistent way.
As these basis sets only have an extra $d$ function compared to cc-pV$n$Z, 
they represent a potential cost savings of 12 basis functions per second-row
atom compared to regular W2 theory. We have considered a minor variant on the
latter (which we term W2w theory), in which aug$'$-cc-pV($n$+d)Z basis sets
are used throughout instead of aug$'$-cc-pV$n$Z+2d1f. (For the geometry optimizations,
cc-pV(T+$d$)Z and cc-pV(Q+$d$)Z are employed instead of their counterparts.)
A comparison with regular W2 theory can be found in the Supplementary Material\cite{EPAPS}:
the two methods perform equivalently, and individual discrepancies for second-row
molecules are very small. 

\section{Definition of W3 theory; attempted definitions of W4 theory}

W3 theory is intended to yield the greatest possible improvement over W2 and W2w theory
at the lowest cost possible. Relative to W2w theory, the following changes
are introduced:
\begin{itemize}
\item the new Douglas-Kroll based scalar relativistic correction was introduced
\item the effect of iterative $T_3$ excitations was estimated from the 
CCSDT--CCSD(T) difference with cc-pVDZ and cc-pVTZ basis sets, then extrapolated
as $a+bL^{-3}$
\item the effect of connected quadruple excitations was estimated as the
CCSDTQ-CCSDT difference with the cc-pVDZ basis set, scaled by a factor of 1.2532 
derived by least-squares fitting to the best available $T_4$ limits over our
training set of molecules
\end{itemize}
We additionally considered two minor modifications. In the first --- denoted W3A theory in this 
paper --- the $T_4$ contribution is computed at the CCSDTQ/ANO431 
level and scaled by 1.275 (scale factor obtained in same manner).  In the second --- denoted
W3K theory in this paper --- the CCSD valence correlation extrapolation is carried out
separately on `singlet' and `triplet' pair correlation energies, as originally advocated by
Klopper\cite{KlopperExtrap} (hence the acronym).

In addition, we considered two attempts at a W4 method, which we will denote here
as W4a and W4b. Relative to W3 theory, the following changes are introduced:
\begin{itemize}
\item the higher-order $T_3$ effect is instead extrapolated from cc-pVTZ and 
cc-pVQZ basis sets
\item in W4a theory, the $T_4$ contribution is computed in the cc-pVTZ basis
set and scaled by 1.13, the scale factor being obtained in the same way as
for W3 theory 
\item in W4b theory, the $T_4$ contribution is instead extrapolated from
the CCSDTQ--CCSDT difference with cc-pVDZ and cc-pVTZ basis sets
\item the inner-shell correlation contribution is extrapolated from
CCSD(T)/aug$'$-cc-pwCVTZ and CCSD(T)/aug$'$-cc-pwCVQZ results
\item the SCF and valence CCSD contributions are extrapolated from
AV5Z and AV6Z basis set combinations
\item the valence (T) contribution from AVQZ and AV5Z
basis set combinations
\end{itemize}

\section{Performance of W3 theory}

We have considered the W2-1 dataset for atomization energies, minus the H$_2$ molecule
(for which W2 and W3 are trivially equivalent) and expanded with
the ozone, N$_2$O, and NO$_2$ molecules. In addition, we have considered subsets of the G2-1
and G2-2 testsets for ionization potentials and electron affinities. Unless
indicated otherwise, experimental data are the same as those in the W2 validation
paper\cite{W1followup}. That is, ionization potentials and electron affinities
were generally taken from the latest edition of the WebBook\cite{webbook}, while with one exception,
atomization energies {\em viz.} heats of formation were critically compiled from
a variety of sources in Ref.\cite{W1followup}. (The exception is the CH diatomic radical, for
which a recent exhaustive computational study\cite{CHnew} has shown that the accepted dissociation 
energy is too low by 0.16 kcal/mol.)

It was previously shown\cite{W1book} that for W2 theory, the use of anharmonic
zero-point energies noticeably improves the mean absolute error: this will be
true {\em a fortiori} for W3 theory. All such ZPVEs were taken from Ref.\cite{W1} 
except for two: ozone (vide infra) and ammonia. For this latter molecule, 
a zero-point energy that properly accounts for the umbrella mode has very recently become available
from the work of Halonen and coworkers\cite{HalonenNH3}: the value of 21.165 kcal/mol
is slightly smaller than the 21.33 kcal/mol computed from the Martin, Lee, and Taylor\cite{Mar92NH3}
quartic force field, used in our previous work.

\subsection{Ionization potentials}

Performance of W2 theory for ionization potentials was quite good
already, and this property is fairly easy to reproduce computationally
in any case. As can be seen in Table \ref{tab:IP},
W3 theory achieves the most significant improvements for CN,
CH$_2$ and for N$_2$, reflecting differential static
correlation contributions in these systems that W3 is better able to cope with. Results
for CO and CS are likewise almost spot-on. Molecules already treated well by 
W2 are likewise treated well by W3. P$_2$ and NH$_2$ display significant 
differences from experiment at the W2 as well as W3 levels, suggesting that the
experimental values may be considerably less reliable than their stated uncertainty.
The WebBook lists a plethora of alternate experimental data for these molecules,
spanning a wide range. 

Performance for the atomic IPs, which are very precisely known experimentally,
is quite satisfying for W3 theory, 
although performance for second-row elements is clearly inferior to that for the
first row. We have considered extrapolations from larger basis sets, post-CCSD(T)
valence correlation contributions extrapolated from the largest basis sets
available (AVTZ and AVQZ), core-valence correlation contributions using larger
basis sets,... and found no significant improvement. One effect we are unable
to cover are post-CCSD(T) contributions to the core-valence correlation, which
would be much more important for second-row than for first-row atoms as both the
core-valence gap is smaller and there are more subvalence electrons.

In all, we can say that W3 theory ought to reliable to 0.01 eV or better.

\subsection{Electron affinities}

Electron affinities are notoriously sensitive to the level of theory (e.g.\cite{ea}),
both in terms of the basis set (as the spatial extent of the wave function differs greatly
between the anion and the parent neutral species) and of the electron correlation method
(as effectively the number of particles is increased). It is in particular well known that
calculating EAs requires the addition of diffuse functions to the basis set\cite{Radom77,Kendall92}. 
Therefore, unmodified W3 theory would fare rather poorly, and we have instead used (diffuse function)
augmented basis sets in the $T_4$ and higher-order $T_3$ corrections. (Regular basis sets were still
used on hydrogen.) Sticklers for acronyms might
prefer to call this approach "W3+ theory".

Not surprisingly (Table \ref{tab:EA}), W3 theory is seen to represent a significant improvement
over W2 theory for this property. W3 results are almost across the board within the
experimental error bar. In fact, our calculations suggest that W3 theory
ought to be competitive with all but the most precise experimental techniques.

\subsection{Molecular total atomization energies}

For molecular atomization energies (Table \ref{tab:TAE}), 
the most spectacular improvement
is seen for the ozone molecule. (Both an accurate re-measurement of
the heat of formation\cite{Taniguchi} and an accurate set of anharmonic
spectroscopic constants\cite{O3anhar} have been published very recently.
As connected quadruple excitations contribute very significantly to
the spectroscopic constants of ozone\cite{BartlettO3T4}, computing
an accurate anharmonic zero-point energy in a large basis set is an
arduous task on which we preferred not to embark for this paper.)
Ozone was omitted from the original W2-1 dataset because of its intrinsic
multireference character: an error of 3 kcal/mol by a method (W2)
that essentially estimates the CCSD(T) limit is not surprising for
a molecule well outside the 'safety envelope' of CCSD(T). W3 theory,
in contrast, puts in a quite respectable performance, with an error
of only 0.38 kcal/mol. 

Very satisfying improvements are likewise seen for two other molecules (N$_2$O
and NO$_2$) with moderate and strong nondynamical correlation effects,
respectively. The W2 errors of 1.20 and 1.16 kcal/mol\cite{W1followup} 
are reduced to 0.51 and 0.09 kcal/mol, respectively.

For some diatomic molecules with
precisely known experimental atomization energies and significant
static correlation, such as F$_2$, O$_2$, NO, and N$_2$, W2 exhibits errors
in the 0.5 kcal/mol range, while W3 reproduces their
dissociation energies basically spot-on. A similar improvement is
seen for the HNO molecule.

In well-behaved systems where W2 performed very well (HF, H$_2$O), so
does W3. It thus satisfies the 'above all, do no harm' requirement. The
mean absolute errors approach the average uncertainty for the experimental data, 0.15 kcal/mol.

Peculiarly, the most significant errors left now are with sulfur systems,
particularly SO$_2$ and H$_2$S. 

Does W3A represent an improvement? Clearly the errors for systems with highly
polar bonds are noticeably reduced, and overall error statistics come down
somewhat. Almost as important, the mean signed error is reduced to near zero.
However, the somewhat marginal reduction in the overall error statistics does
not appear to justify the substantially increased computational cost (factor of about 4--5,
dominated by the $T_4$ step). More fundamentally, the increase in the number of CCSDTQ amplitudes
by about the same factor may easily make the difference between a calculation that is just
feasible with available hardware and one that is not. For systems with strongly polar bonds, 
W3A, if practically feasible, may serve as an additional check on a W3 prediction. 

The added cost of W3K over W3, by contrast, is nil in open-shell cases and quite
modest in closed-shell cases\cite{closedMOLPRO}. Table \ref{tab:TAE} reveals that W3K
represents a marginal overall improvement over standard W3. However, its performance for
second-row systems is markedly superior, and in this sense it is arguably  a more `balanced' method
than standard W3. For first-row systems, reduced deviations for systems dominated by dynamical
correlation are offset by increased deviations for systems with multireference character.
The choice between W3 and W3K can be argued either way, and we have simply left the choice
open to the user.

\section{Performance of W4a and W4b theory. Outlook for further improvements.}

Some of the systems were small enough that we could compute W4a and W4b
total atomization energies. A comparison is given in Table \ref{tab:W4}.
First of all, W4a (with its scaling-based $T_4$ correction) is clearly
superior to W4b (with its extrapolation-based $T_4$ correction). The extrapolation
misbehaves in O and F molecules, as the $T_4$ correction appears to go
through a minimum as a function of the basis set for PVTZ.
Secondly, despite the formidable added computational cost, overall
performance of W4a only represents a marginal improvement over W3.

This begs the question as to what is still missing in W4a and W4b theory. Three factors
suggest themselves:

\noindent (a) $T_5$ effects (vide supra). These will primarily affect
systems with strong nondynamical correlation effects, and at least some
of the systems where W4a and W4b `cannot make the grade' are essentially
devoid of these.

\noindent (b) nonadiabatic effects. Literature values for DBOC (diagonal 
Born-Oppenheimer Corrections) are available for some hydrogen-containing systems\cite{DBOC}:
SH 0.2 cm$^{-1}$, i.e. essentially nil for our purposes; CH$_2$($^3B_1$) +0.05 kcal/mol; 
CH radical -0.05 kcal/mol; OH radical -0.01 kcal/mol; H$_2$O +0.10 kcal/mol;
HF -0.04 kcal/mol. For the all-heavy
atom systems we can safely consider the DBOC to be negligible on the scale of interest to
us. Taking DBOCs into account may thus somewhat improve results for some hydrides.

\noindent (c) post-CCSD(T) effects in the core-valence correlation
contribution. Explicit calculation of such effects is an arduous 
task, but all-electron CCSDT calculations on N$_2$ and B$_2$
suggest contributions on the order of 0.05--0.10 kcal/mol.
(For B$_2$, we additionally found a $T_4$ core-valence contribution to
the dissociation energy of 0.04 kcal/mol.)
For second-row molecules, with smaller core-valence gaps and
more subvalence electrons, this contribution is liable to 
be more important: this is consistent with our general observation
that W3, W4a, and W4b theory all perform significantly better
for first-row than for second-row systems.

\noindent (d) higher-order relativistic effects.
Second-order spin-order coupling was found\cite{Feller2003} to contribute 
2 kcal/mol to the binding energy of I$_2$ and 0.4 kcal/mol to that of Br$_2$;
it cannot be ruled out that the contribution for Cl$_2$ would reach 0.1 kcal/mol.
Recently, the Lamb shift was found\cite{Dyall2001} to contribute +0.04 and +0.07 kcal/mol, 
respectively, to the binding energy of BF$_3$ and AlF$_3$.

\noindent (e) finally, although the total energy depends fairly weakly 
on geometric 
displacements near the equilibrium geometry, the small discrepancies between
CCSD(T)/cc-pV(Q+d)Z and exact bottom-of-the-well 
reference geometries may cause small errors. This, however, 
clearly cannot explain the issues we are having with atomic IPs and EAs.

\section{Conclusions}

We have developed and validated a new computational thermochemistry protocol termed
W3 theory. Compared to the older W2 theory\cite{W1}, the main improvements are 
an improved treatment of scalar relativistic effects, and particularly an approximate 
account for post-CCSD(T) correlation effects. The new method is appreciably more costly,
but considerably more robust, than W2 theory, and in particular yields reliable results
for systems suffering from significant nondynamical correlation effects. It confirms
the earlier assertion\cite{W1followup} that the accuracy of W2 theory is basically limited
by that of the CCSD(T) method.

Iterative $T_3$ effects exhibit a basis set convergence behavior similar to the $T_3$
contribution overall. They almost universally decrease molecular binding energies.
Included by themselves, they yield less accurate results than 
CCSD(T) almost across the board: it is only when $T_4$ effects are included that superior
performance is achieved. $T_4$ effects systematically increase molecular binding energies.
Their basis set convergence is quite rapid, and even CCSDTQ/cc-pVDZ scaled by 1.2532 will
yield a quite passable quadruples contribution. The effect of still higher-order excitations 
was gauged for a subset of molecules (notably the eight-valence electron systems): $T_5$
contributions reach 0.3 kcal/mol for the pathologically multireference $X~^1\Sigma^+_g$ state
of C$_2$ but are quite small for other systems.

Over a sample of 30 molecules --- including some with severe nondynamical correlation 
effects --- going from W2 to W3 reduces mean absolute error in total atomization energies from
0.395 to 0.220 kcal/mol, RMS error from 0.696 to 0.280 kcal/mol, and the two largest individual 
errors from \{+3.0 (O$_3$), +1.2 (N$_2$O, NO$_2$)\} kcal/mol to \{-0.78 (SO$_2$), +0.51 kcal/mol (N$_2$O)\}.

Various avenues for further enhancing the accuracy of W3 theory were explored, including 
more extensive basis sets, BSSE corrections, larger-basis set corrections for $T_4$ and 
higher-order $T_3$ effects, and extrapolation of the inner-shell correlation effects to the
basis set limit. Only marginal improvements can be achieved by these costly measures: W3 
appears to be `scratching the bottom out of the barrel'. BSSE on molecular binding energies
is still significant even with basis sets as large as the AV6Z combination, but is almost entirely
removed by the extrapolation. We speculate that the main obstacle
to breaking the 0.1 kcal/mol barrier would be CCSD(T) imperfections in the core-valence
correlation energy; their explicit computation is presently impractical for all but the very
smallest systems. Lesser potential error sources include, but are not limited to,
post-CCSDTQ valence correlation effects, corrections to the Born-Oppenheimer approximation,
higher-order relativistic effects (second-order spin-orbit coupling, Lamb shift,...) and 
imperfections in the reference geometry.

\begin{acknowledgments}
JMLM is a member of the Lise Meitner-Minerva Center for Computational
Quantum Chemistry. ADB acknowledges a postdoctoral fellowship from
the Feinberg Graduate School (Weizmann Institute). Research at Weizmann
was supported by the Minerva Foundation, Munich, Germany, and by the
Helen and Martin Kimmel Center for Molecular Design. JG acknowledges support
from the Fonds der Chemischen Industrie (Germany).
\end{acknowledgments}

\clearpage

\begingroup
\squeezetable
\begin{table}
\caption{Basis set convergence of $T_4$ and higher-order $T_3$ effects on atomic
ionization potentials and electron affinities\label{tab:T4IP}}

\begin{tabular}{lrrrrrrrr}
\hline\hline

           &             \multicolumn{ 4}{c}{CCSDT - CCSD(T) } &               \multicolumn{ 4}{c}{CCSDTQ - CCSDT} \\

           &       AVDZ &       AVTZ &       AVQZ &      limit$^a$ &       AVDZ &       AVTZ &       AVQZ &      limit$^a$ \\
\hline
                                                                  \multicolumn{ 9}{c}{Effect on ionization potentials (meV)} \\
\hline

         B &      11.24 &      11.97 &      11.05 &      10.38 &       0.00 &       0.00 &       0.00 &       0.00 \\

         C &       3.86 &       5.08 &       4.58 &       4.21 &       1.09 &       1.30 &       1.41 &       1.49 \\

         N &       0.22 &       1.10 &       0.85 &       0.66 &       0.55 &       0.70 &       0.86 &       0.98 \\

         O &       4.00 &       3.53 &       3.08 &       2.76 &       1.30 &       1.08 &       1.48 &       1.77 \\

         F &       1.71 &       0.54 &      -0.11 &      -0.59 &       1.71 &       1.14 &       1.76 &       2.21 \\

        Ne &      -0.03 &      -3.02 &      -3.94 &      -4.61 &       2.24 &       0.90 &       1.72 &       2.31 \\

        Al &      11.97 &      13.69 &      12.45 &      11.54 &       0.00 &       0.00 &       0.00 &       0.00 \\

        Si &       4.32 &       9.27 &       8.08 &       7.22 &       1.74 &       2.34 &       2.61 &       2.80 \\

         P &      -0.23 &       3.54 &       3.05 &       2.70 &       2.24 &       2.35 &       2.97 &       3.42 \\

         S &       4.74 &       2.51 &       3.03 &       3.40 &       1.76 &       1.66 &       2.40 &       2.93 \\

        Cl &       0.58 &      -0.58 &      -0.83 &      -1.01 &       1.93 &       2.32 &       2.88 &       3.28 \\

        Ar &      -2.21 &      -4.77 &      -5.81 &      -6.56 &       2.48 &       2.63 &       3.16 &       3.56 \\
\hline
                                                                    \multicolumn{ 9}{c}{Effect on electron affinities (meV)} \\
\hline
         B &      14.87 &      14.98 &      14.06 &      13.39 &       3.73 &       4.76 &       5.01 &       5.19 \\

         C &       9.52 &       9.18 &       8.26 &       7.59 &       4.20 &       4.82 &       5.15 &       5.39 \\

         O &      10.28 &       5.11 &       3.29 &       1.96 &       9.97 &       8.98 &      10.54 &      11.67 \\

         F &       3.32 &      -5.52 &      -8.39 &     -10.48 &      10.34 &       6.89 &       8.68 &       9.98 \\

        Al &       9.85 &      12.01 &      10.42 &       9.26 &       3.07 &       4.20 &       4.55 &       4.81 \\

        Si &       4.60 &       6.04 &       4.08 &       2.64 &       3.84 &       4.79 &       4.72 &       4.67 \\

         P &      12.80 &       7.30 &       7.00 &       6.78 &       4.07 &       5.90 &       6.85 &       7.55 \\

         S &       4.92 &       1.27 &       0.15 &      -0.67 &       4.28 &       6.38 &       7.42 &       8.18 \\

        Cl &      -1.13 &      -5.61 &      -7.30 &      -8.53 &       4.41 &       5.75 &       6.94 &       7.82 \\
\hline\hline
\end{tabular}  

(a) From AVQZ + (AVQZ -- AVTZ)/($(4/3)^3-1$) \cite{Halkier}

\end{table}
\endgroup

\begingroup
\squeezetable
\begin{table}
\caption{Basis set convergence of $T_4$ effects on molecular
total atomization energies (kcal/mol)\label{tab:T4TAE}}

\begin{tabular}{lrrrrrrr}
\hline\hline
           &       PVDZ &       AVDZ &       PVTZ &       AVTZ &  \{PVDZ,PVTZ\}$^a$ & \{AVDZ,AVTZ\}$^a$ & ANO431 \\
\hline
       H$_2$O &       0.24 &       0.25 &       0.17 &            &       0.14 &            & 0.17 \\

        B$_2$ &       0.99 &       1.03 &       1.19 &       1.21 &       1.26 &       1.27 & 1.02 \\

      C$_2$H$_2$ &       0.54 &       0.58 &            &            &            &            & 0.53 \\

       CH$_3$ &       0.06 &       0.06 &       0.05 &            &       0.05 &            & 0.04 \\

       CH$_4$ &       0.07 &       0.07 &            &            &            &            & 0.05 \\

        CH &       0.03 &       0.03 &       0.03 &       0.03 &       0.03 &       0.03 & 0.03 \\

       CO$_2$ &       0.99 &            &            &            &            &            & 0.84 \\

        CO &       0.53 &       0.59 &       0.56 &            &       0.56 &            & 0.47 \\

        F$_2$ &       0.82 &       0.98 &       0.80 &            &       0.79 &            & 0.73 \\

        HF &       0.17 &       0.17 &       0.09 &       0.10 &       0.06 &       0.07 & 0.11 \\

        N$_2$ &       0.87 &       0.96 &       0.94 &            &       0.96 &            & 0.86 \\

       NH$_3$ &       0.17 &       0.19 &       0.15 &            &            &            & 0.15 \\

       NNO &       1.75 &            &            &            &            &            & 1.67  \\

        NO &       0.75 &       0.84 &       0.78 &            &       0.79 &            & 0.69 \\

        O$_2$ &       1.08 &       1.19 &       1.07 &            &       1.07 &            & 0.99 \\

        O$_3$ &       3.21 &            &            &            &            &            & 3.17 \\

        C$_2$ &       1.59 &       1.77 &       2.12 &            &       2.31 &            & 1.71 \\

        BN &       1.38 &       1.56 &       1.87 &            &       2.05 &            & 1.48 \\

       MgO &       1.55 &       1.54 &       1.74 &       1.69 &       1.81 &       1.75 & 1.37 \\

       BeO &       0.69 &       0.68 &       0.67 &            &       0.66 &            & 0.51 \\

        CN &       0.84 &       0.92 &       0.99 &            &       1.05 &            & 0.84 \\

       NO$_2$ &       1.71 &            &            &            &            &            & 1.61  \\

       Cl$_2$ &       0.24 &            &       0.39 &            &       0.45 &            & 0.24 \\

       ClF &       0.39 &            &       0.41 &            &       0.42 &            & 0.31 \\

        CS &       0.50 &            &       0.87 &            &       1.00 &            & 0.56 \\

       H$_2$S &       0.08 &            &       0.13 &            &       0.15 &            & 0.07 \\

       HCl &       0.06 &            &       0.09 &            &       0.10 &            & 0.06 \\

      HOCl &       0.48 &            &            &            &            &            & 0.41 \\

       PH$_3$ &       0.05 &            &       0.09 &            &       0.10 &            & 0.04 \\

        SO &       0.73 &            &       0.79 &            &       0.82 &            & 0.63 \\
        SO$_2$ & 1.44\\
        OCS    & 0.98\\
    ClCN       & 0.94\\
    C$_2$H$_4$ & 0.33 & & & & & & 0.30\\
    H$_2$CO    & 0.50 & & & & & & 0.42\\
    HNO        & 0.65 & & & & & & 0.60\\
\hline\hline
\end{tabular}  

(a) extrapolated from the two basis sets indicated

\end{table}
\endgroup

\begingroup
\squeezetable
\begin{table}
\caption{Basis set convergence of higher-order $T_3$ effects on molecular
total atomization energies (kcal/mol)\label{tab:T3TAE}}

\begin{tabular}{lrrrrrrrr}
\hline\hline
           &       PVDZ &       AVDZ &       PVTZ &       AVTZ &  \{PVDZ,PVTZ\}$^a$ & \{AVDZ,AVTZ\}$^a$ &  PVQZ & \{PVTZ,PVQZ\} \\
\hline
       H$_2$O &       0.04 &      -0.02 &      -0.11 &      -0.18 &      -0.17 &      -0.23 &      -0.16 &      -0.18 \\

        B$_2$ &       0.58 &       0.54 &       0.30 &       0.24 &       0.19 &       0.13 &       0.17 &       0.12 \\

      C$_2$H$_2$ &      -0.12 &      -0.25 &      -0.51 &      -0.62 &      -0.66 &      -0.75 &      -0.61 &      -0.64 \\

      C$_2$H$_4$ &       0.03 &      -0.08 &      -0.28 &      -0.38 &      -0.40 &      -0.49 &            &            \\

       CH$_3$ &       0.06 &       0.05 &       0.01 &      -0.02 &      -0.01 &      -0.05 &            &            \\

       CH$_4$ &       0.06 &       0.04 &      -0.03 &      -0.06 &      -0.06 &      -0.10 &            &            \\

        CH &       0.13 &       0.14 &       0.12 &       0.10 &       0.12 &       0.09 &       0.12 &       0.12 \\

       CO$_2$ &      -0.14 &      -0.48 &      -0.72 &      -0.93 &      -0.94 &      -1.10 &      -0.88 &      -0.93 \\

        CO &       0.05 &      -0.12 &      -0.35 &      -0.46 &      -0.49 &      -0.59 &      -0.44 &      -0.48 \\

        F$_2$ &       0.08 &      -0.05 &      -0.21 &      -0.27 &      -0.32 &      -0.35 &      -0.26 &      -0.27 \\

      H$_2$CO &       0.05 &      -0.09 &      -0.32 &      -0.44 &      -0.46 &      -0.57 &            &            \\

        HF &       0.01 &      -0.01 &      -0.09 &      -0.11 &      -0.12 &      -0.15 &      -0.12 &      -0.13 \\

       HNO &       0.43 &       0.26 &       0.12 &      -0.03 &       0.00 &      -0.13 &            &            \\

        N$_2$ &      -0.05 &      -0.23 &      -0.50 &      -0.67 &      -0.66 &      -0.84 &      -0.59 &      -0.63 \\

       NH$_3$ &       0.12 &       0.07 &      -0.03 &      -0.11 &      -0.08 &      -0.17 &            &            \\

       NNO &      -0.41 &      -0.77 &      -1.10 &      -1.37 &      -1.35 &      -1.59 &            &            \\

        NO &       0.13 &      -0.05 &      -0.31 &      -0.45 &      -0.47 &      -0.60 &      -0.40 &      -0.44 \\

        O$_2$ &      -0.06 &      -0.26 &      -0.52 &      -0.64 &      -0.68 &      -0.78 &      -0.63 &      -0.67 \\

        O$_3$ &      -0.10 &      -0.77 &      -0.92 &      -1.28 &      -1.23 &      -1.47 &            &            \\

        C$_2$ &      -1.22 &      -1.48 &      -1.87 &      -2.02 &      -2.12 &      -2.22 &      -2.06 &      -2.13 \\

        BN &      -1.95 &      -2.07 &      -2.40 &      -2.51 &      -2.57 &      -2.68 &      -2.50 &      -2.54 \\

       MgO &      -0.01 &      -0.21 &      -0.64 &      -0.78 &      -0.87 &      -0.99 &            &            \\

       BeO &       0.58 &       0.39 &       0.04 &      -0.06 &      -0.16 &      -0.22 &            &            \\

        CN &       0.41 &            &      -0.08 &            &      -0.26 &            &      -0.19 &      -0.23 \\

       NO$_2$ &       0.04 &            &      -0.68 &            &      -0.95 &            &            &            \\

       Cl$_2$ &       0.02 &            &      -0.25 &            &      -0.35 &            &      -0.33 &      -0.36 \\

       ClF &       0.05 &            &      -0.19 &            &      -0.28 &            &      -0.24 &      -0.26 \\

        CS &       0.11 &            &      -0.39 &            &      -0.57 &            &      -0.50 &      -0.55 \\

       H$_2$S &       0.09 &            &      -0.04 &            &      -0.09 &            &      -0.07 &      -0.08 \\

       HCl &       0.02 &            &      -0.07 &            &      -0.10 &            &      -0.10 &      -0.11 \\

      HOCl &       0.03 &            &      -0.31 &            &      -0.43 &            &       0.00 &       0.00 \\

       PH$_3$ &       0.20 &            &       0.04 &            &      -0.02 &            &       0.05 &       0.05 \\

        SO &      -0.06 &            &      -0.55 &            &      -0.74 &            &      -0.66 &      -0.70 \\

       SO$_2$ &      -0.10 &            &      -0.90 &            &      -1.19 &            &            &            \\

       OCS &      -0.09 &            &      -0.76 &            &      -1.02 &            &            &            \\

      CNCl &      -0.22 &            &      -0.83 &            &      -1.05 &            &            &            \\
\hline\hline
\end{tabular}  

(a) extrapolated from the two basis sets indicated

\end{table}
\endgroup

\begingroup
\squeezetable
\begin{table}
\caption{Comparison of scalar relativistic corrections for molecular total atomization energies (kcal/mol)\label{tab:Mikhal}}
\begin{center}
\begin{tabular}{lcccc}
\hline\hline
Molecule & ACPF/MTsmall$^a$ & CCSD(T)/ARVQZ &\{ARVTZ,ARVQZ\}$^b$ & \{ARVQZ,ARV5Z\}$^b$ \\
\hline
H$_2$          & $\phantom{-}$0.00  &$-$0.001 & $-$0.001 & $-$0.001 \\
N$_2$          & $-$0.11 &$-$0.133 & $-$0.146 & $-$0.145 \\
O$_2$          & $-$0.15 &$-$0.176& $-$0.184 & $-$0.191 \\
F$_2$           & $+$0.03 &$-$0.024& $-$0.033 & $-$0.034 \\
HF                 & $-$0.20 &$-$0.194& $-$0.196 & $-$0.198 \\
CH                & $-$0.03 &$-$0.040& $-$0.041 & $-$0.039 \\
CO                & $-$0.14 &$-$0.157& $-$0.166 & $-$0.162 \\
NO                & $-$0.16 &$-$0.185& $-$0.193 & $-$0.194 \\
CS                 & $-$0.15 &$-$0.159& $-$0.141 & $-$0.140 \\
SO                 & $-$0.31 &$-$0.336& $-$0.344 & $-$0.353 \\
HCl                & $-$0.26 &$-$0.246& $-$0.249 & $-$0.239 \\
ClF                & $-$0.12 &$-$0.177& $-$0.205 & $-$0.172 \\
Cl$_2$         & $-$0.15 &$-$0.208& $-$0.242 & $-$0.190 \\
HNO             & $-$0.24 &$-$0.266& $-$0.274 & $-$0.274 \\
CO$_2$       & $-$0.45 &$-$0.471& $-$0.486 & $-$0.477 \\
H$_2$O       & $-$0.26 &$-$0.264& $-$0.268 & $-$0.269 \\
H$_2$S        & $-$0.41 &$-$0.393& $-$0.400 & $-$0.399 \\
HOCl             & $-$0.28 &$-$0.323& $-$0.340 & $-$0.325 \\
OCS               & $-$0.53 &$-$0.530& $-$0.547 & $-$0.542 \\
ClCN              & $-$0.43 &$-$0.442& $-$0.451 & $-$0.446 \\
SO$_2$         & $-$0.71 &$-$0.814& $-$0.837 & $-$0.857 \\
CH$_3$         & $-$0.17 &$-$0.172& $-$0.173 & $-$0.168 \\
NH$_3$         & $-$0.25 &$-$0.251& $-$0.245 & $-$0.243 \\
PH$_3$          & $-$0.46 &$-$0.453& $-$0.460 & $-$0.455 \\
C$_2$H$_2$ & $-$0.27 &$-$0.280& $-$0.287 & $-$0.270 \\
CH$_2$O       & $-$0.32 &$-$0.334& $-$0.340 & $-$0.335 \\
CH$_4$          & $-$0.19 &$-$0.193& $-$0.195 & $-$0.187 \\
C$_2$H$_4$ & $-$0.33 &$-$0.332& $-$0.336 & $-$0.324 \\
SiF$_4$           & $-$1.88 & --- & $-$1.895 & --- \\
SO$_3$           & $-$1.71 & --- & $-$1.829 & $-$1.878 \\
\hline
\multicolumn{2}{l}{Mean Absolute Deviation} & 0.03 & 0.03 \\
\multicolumn{2}{l}{MAD without SO$_2$, SO$_3$} & 0.02 & 0.02 \\
\hline\hline
\end{tabular}

(a) \footnotesize{Data taken from~\cite{W1book}, except SiF$_4$~\cite{Martin1999a}
and SO$_3$~\cite{Martin1999b}.}

(b) extrapolated from the two basis sets indicated

\end{center}

\end{table}
\endgroup

\begingroup
\squeezetable
\begin{table}
\caption{Comparison of different extrapolation procedures for the SCF and 
valence correlation energy (kcal/mol)\label{tab:extrap}$^a$}
\begin{tabular}{l|c|ccccc|c}
\hline\hline
 & SCF & \multicolumn{5}{c}{CCSD -- SCF} & (T) \\
Basis sets &\{AV5Z,AV6Z\}&\multicolumn{2}{c}{\{AVTZ,AVQZ\}}&\{AVQZ,AV5Z\}&\multicolumn{2}{c}{\{AV5Z,AV6Z\}}&\{AVQZ,AV5Z\}\\
Extrap.$^c$ & 5    & 3    & 3,5  & 3,5  & 3    & 3,5  & 3    \\
\hline
C$_2$H$_2$  & -0.032  &  0.292  &  0.259  & -0.104  & -0.022  & -0.071  & -0.025 \\
CH          & -0.003  &  0.077  &  0.145  &  0.000  & -0.014  & -0.015  & -0.009 \\
CH$_3$      & -0.022  &  0.258  &  0.341  & -0.025  & -0.030  & -0.042  & -0.022 \\
CH$_4$      & -0.029  &  0.259  &  0.315  & -0.048  & -0.031  & -0.055  & -0.030 \\
CO$_2$      &  0.017  & -0.003  & -0.236  & -0.216  &  0.003  & -0.093  &   \\
H$_2$O      &  0.000  &  0.282  &  0.244  & -0.077  & -0.085  & -0.120  & -0.023 \\
HF          &  0.007  &  0.293  &  0.215  & -0.060  & -0.015  & -0.043  & -0.017 \\
N$_2$O      &  0.003  &  0.014  & -0.071  & -0.169  &  0.121  &  0.042  &   \\
NO          &  0.013  & -0.148  & -0.166  & -0.089  &  0.021  & -0.019  &  0.036 \\
O$_2$       &  0.014  &  0.014  & -0.079  & -0.080  &  0.059  &  0.031  &  0.030 \\
N$_2$       & -0.001  & -0.287  & -0.216  & -0.086  & -0.079  & -0.117  &  0.039 \\
CO          & -0.002  & -0.103  & -0.228  & -0.127  & -0.022  & -0.078  &  0.026 \\
F$_2$       & -0.007  &  0.091  & -0.177  & -0.128  &  0.051  & -0.009  &  0.012 \\
Cl$_2$      & -0.012  & -0.442  & -0.729  & -0.180  & -0.306  & -0.381  &  0.056 \\
ClF         &  0.044  & -0.124  & -0.391  & -0.146  & -0.100  & -0.164  &  0.021 \\
CS          &  0.038  & -0.386  & -0.467  & -0.113  & -0.243  & -0.285  &  0.083 \\
H$_2$S      &  0.028  &  0.192  &  0.178  & -0.066  & -0.122  & -0.146  &  0.019 \\
HCl         &  0.003  &  0.053  & -0.005  & -0.060  & -0.133  & -0.158  &  0.014 \\
HOCl        &  0.020  & -0.034  & -0.255  & -0.162  & -0.154  & -0.226  &  0.010 \\
PH$_3$      &  0.052  &  0.441  &  0.587  & -0.025  & -0.069  & -0.076  &  0.026 \\
SO          &  0.053  & -0.298  & -0.439  & -0.103  & -0.121  & -0.156  &  0.032 \\
SO$_2$ (d)  &  0.176  & -0.629  & -1.063  & -0.320  & -0.217  & -0.352  &  0.056 \\
OCS         &  0.020  & -0.011  & -0.243  & -0.203  & -0.179  & -0.263  &   \\
NH$_3$      & -0.011  &  0.262  &  0.382  & -0.037  & -0.105  & -0.124  &  0.026 \\
\hline\hline
\end{tabular}

(a) All values relative to the standard W2 procedures.

(b) extrapolated from the two basis sets indicated

(c) "3,5" indicates separate extrapolation of singlet-coupled pairs by $E(L)=E_\infty+a/L^3$ and of
triplet pairs by $E(L)=E_\infty+a/L^5$; "3" a joint extrapolation by $E(L)=E_\infty+a/L^3$ ; and similarly for "5"

(d) SCF/aug-cc-pV6Z+2d1f: 121.94 kcal/mol. 3-point geometric extrapolation: aug-cc-pV(X+d)Z (X=Q,5,6): 121.95 kcal/mol;
aug-cc-pVXZ+2d1f: 121.91 kcal/mol. Best estimate: 121.93$\pm$0.04 kcal/mol.
\end{table}
\endgroup

\begingroup
\squeezetable
\begin{table}
\caption{Effect of basis set superposition error aon raw and extrapolated valence correlation energies (kcal/mol)\label{tab:BSSE}}
\begin{tabular}{lrrrr}
\hline\hline
           &
       BSSE &
       BSSE &
  BSSE CCSD &
  BSSE CCSD \tabularnewline
\hline
  molecule &
 \{AVQZ,AV5Z\}$^a$ &
 \{AV5Z,AV6Z\}$^a$ &
   AV5Z &
   AV6Z \tabularnewline
\hline
       CH$_4$ &
      0.071 &
      0.018 &
      0.217 &
      0.120 \tabularnewline
      C$_2$H$_2$ &
      0.127 &
      0.026 &
      0.330 &
      0.183 \tabularnewline
       CH$_3$ &
      0.050 &
      0.012 &
      0.197 &
      0.110 \tabularnewline
        CH &
      0.012 &
      0.006 &
      0.073 &
      0.040 \tabularnewline
       NH$_3$ &
      0.057 &
      0.026 &
      0.266 &
      0.145 \tabularnewline
       H$_2$O &
      0.015 &
      0.021 &
      0.359 &
      0.200 \tabularnewline
        HF &
      0.017 &
      0.007 &
      0.283 &
      0.161 \tabularnewline
        O$_2$ &
      0.128 &
      0.066 &
      0.472 &
      0.246 \tabularnewline
        NO &
      0.119 &
      0.052 &
      0.403 &
      0.213 \tabularnewline
        N$_2$ &
      0.112 &
      0.047 &
      0.295 &
      0.153 \tabularnewline
        CO &
      0.103 &
      0.053 &
      0.403 &
      0.212 \tabularnewline
        F$_2$ &
      0.115 &
      0.042 &
      0.293 &
      0.151 \tabularnewline
       Cl$_2$ &
     -0.160 &
      0.101 &
      0.368 &
      0.165 \tabularnewline
       ClF &
     -0.015 &
      0.081 &
      0.392 &
      0.189 \tabularnewline
        CS &
     -0.023 &
      0.057 &
      0.376 &
      0.191 \tabularnewline
       H$_2$S &
     -0.069 &
      0.074 &
      0.328 &
      0.181 \tabularnewline
       HCl &
     -0.119 &
      0.032 &
      0.307 &
      0.162 \tabularnewline
      HOCl &
     -0.023 &
      0.085 &
      0.385 &
      0.184 \tabularnewline
       PH$_3$ &
     -0.021 &
      0.032 &
      0.183 &
      0.106 \tabularnewline
        SO &
      0.060 &
      0.067 &
      0.461 &
      0.238 \tabularnewline
       SO$_2$ &
      0.134 &
      0.122 &
      0.811 &
      0.417 \tabularnewline
\hline\hline
\end{tabular}

(a) extrapolated from the two basis sets indicated

\end{table}
\endgroup

\begingroup
\squeezetable
\begin{table}
\caption{Effect on TAE (kcal/mol) of an improved inner-shell correlation treatment\label{tab:CORE}}
\begin{tabular}{lrrrrrr}
\hline\hline
  molecule &
   aug$'$-cc-pwCVTZ &
   aug$'$-cc-pwCVQZ &
 extrapolated &
  BSSE (TZ) &
  BSSE (QZ) &
 BSSE (extrap.) \tabularnewline
\hline
       CH$_4$ &
       1.12 &
       1.21 &
       1.27 &
       0.06 &
       0.02 &
       0.02 \tabularnewline
       NH$_3$ &
       0.57 &
       0.62 &
       0.65 &
       0.04 &
       0.01 &
       0.01 \tabularnewline
      C$_2$H$_2$ &
       2.16 &
       2.35 &
       2.49 &
       0.11 &
       0.02 &
       0.04 \tabularnewline
       CH$_3$ &
       0.95 &
       1.03 &
       1.09 &
       0.05 &
       0.01 &
       0.01 \tabularnewline
        CH &
       0.13 &
       0.14 &
       0.14 &
       0.01 &
       0.00 &
       0.00 \tabularnewline
       H$_2$O &
       0.34 &
       0.37 &
       0.38 &
       0.02 &
       0.01 &
       0.00 \tabularnewline
        HF &
       0.20 &
       0.18 &
       0.16 &
       0.01 &
       0.00 &
       0.00 \tabularnewline
        O$_2$ &
       0.25 &
       0.24 &
       0.23 &
       0.04 &
       0.01 &
       0.02 \tabularnewline
        NO &
       0.38 &
       0.40 &
       0.41 &
       0.05 &
       0.01 &
       0.02 \tabularnewline
        N$_2$ &
       0.67 &
       0.74 &
       0.79 &
       0.06 &
       0.01 &
       0.02 \tabularnewline
        CO &
       0.82 &
       0.90 &
       0.96 &
       0.06 &
       0.01 &
       0.02 \tabularnewline
        F$_2$ &
      -0.06 &
      -0.08 &
      -0.10 &
       0.02 &
       0.00 &
       0.01 \tabularnewline
       Cl$_2$ &
       0.24 &
       0.18 &
       0.14 &
       0.10 &
       0.05 &
       0.00 \tabularnewline
       ClF &
       0.13 &
       0.08 &
       0.04 &
       0.08 &
       0.03 &
       0.00 \tabularnewline
        CS &
       0.72 &
       0.79 &
       0.84 &
       0.15 &
       0.07 &
      -0.01 \tabularnewline
       H$_2$S &
       0.28 &
       0.31 &
       0.33 &
       0.13 &
       0.08 &
      -0.04 \tabularnewline
       HCl &
       0.17 &
       0.18 &
       0.19 &
       0.06 &
       0.03 &
      -0.02 \tabularnewline
      HOCl &
       0.33 &
       0.29 &
       0.26 &
       0.09 &
       0.04 &
       0.00 \tabularnewline
       PH$_3$ &
       0.25 &
       0.31 &
       0.34 &
       0.20 &
       0.12 &
      -0.06 \tabularnewline
        SO &
       0.47 &
       0.48 &
       0.49 &
       0.15 &
       0.07 &
      -0.01 \tabularnewline
       SO$_2$ &
       0.92 &
       0.95 &
       0.97 &
       0.85 &
       0.33 &
       0.04 \tabularnewline
\hline\hline
\end{tabular}  

\end{table}
\endgroup

\begingroup
\squeezetable
\begin{table}
\caption{Errors (experiment--theory) for computed ionization potentials (eV)\label{tab:IP}}

\begin{tabular}{lrrr}
\hline\hline
  molecule &
         W2 &
         W3 &
 Expt. uncertainty \tabularnewline
\hline
      B &
       0.007 &
      -0.005 &
       0.00002 \tabularnewline
      C &
       0.010 &
       0.004 &
       0.0001 \tabularnewline
      N &
       0.000 &
      -0.002 &
       0.001 \tabularnewline
      O &
       0.005 &
       0.000 &
       0.001 \tabularnewline
      F &
       0.002 &
       0.001 &
       0.001 \tabularnewline
      Ne &
       0.000 &
      -0.002 &
       0.001 \tabularnewline
      Al &
       0.023 &
       0.009 &
       0.001 \tabularnewline
      Si &
       0.018 &
       0.005 &
       0.00003 \tabularnewline
      P &
       0.011 &
       0.005 &
       0.00001 \tabularnewline
      S &
       0.014 &
       0.012 &
       0.001 \tabularnewline
      Cl &
       0.007 &
       0.007 &
       0.001 \tabularnewline
      Ar &
       0.009 &
       0.013 &
       0.001 \tabularnewline
      C$_2$H$_2$ &
     -0.004 &
      0.008 &
      0.001 \tabularnewline
      C$_2$H$_4$ &
     -0.001 &
      0.004 &
      0.000 \tabularnewline
       CH$_2$ &
      0.023 &
      0.010 &
      0.003 \tabularnewline
       CH$_4$ &
     -0.033 &
     -0.030 &
      0.010 \tabularnewline
       Cl$_2$ &
     -0.008 &
      0.005 &
      0.003 \tabularnewline
       ClF &
      0.005 &
      0.018 &
      0.010 \tabularnewline
        CN &
     -0.046 &
     -0.014 &
      0.020 \tabularnewline
        CO &
     -0.014 &
     -0.003 &
      0.000 \tabularnewline
        CS &
     -0.017 &
      0.001 &
      0.010 \tabularnewline
       H$_2$O &
      0.006 &
      0.006 &
      0.000 \tabularnewline
       H$_2$S &
     -0.008 &
     -0.006 &
      0.001 \tabularnewline
        HF &
     -0.016 &
     -0.018 &
      0.003 \tabularnewline
        N$_2$ &
     -0.046 &
      0.000 &
      0.008 \tabularnewline
       NH$_2$ &
     -0.034 &
     -0.038 &
      0.010 \tabularnewline
       NH$_3$ &
     -0.004 &
     -0.004 &
      0.090 \tabularnewline
        NH &
     -0.046 &
     -0.052 &
      0.010 \tabularnewline
        O$_2$ &
     -0.024 &
      0.002 &
      0.000 \tabularnewline
        OH &
      0.001 &
     -0.004 &
      0.000 \tabularnewline
        P$_2$ &
      0.047 &
      0.065 &
      0.002 \tabularnewline
       PH$_2$ &
      0.003 &
      0.000 &
      0.002 \tabularnewline
       PH$_3$ &
     -0.006 &
     -0.012 &
      0.002 \tabularnewline
        PH &
     -0.006 &
     -0.011 &
      0.008 \tabularnewline
        S$_2$ &
     -0.011 &
      0.012 &
      0.002 \tabularnewline
        SH &
      0.007 &
      0.006 &
      0.000 \tabularnewline
      SiH$_4$ &
      0.006 &
      0.006 &
      0.020 \tabularnewline
\hline
  mean abs &
      0.0141 &
      0.0104 &
            \tabularnewline
       RMS &
      0.0202 &
      0.0161 &
            \tabularnewline
    max(+) &
         P$_2$ &
         P$_2$ &
            \tabularnewline
           &
      0.047 &
      0.065 &
            \tabularnewline
    max(-) &
         CN/N$_2$ &
         NH$_2$ &
            \tabularnewline
           &
     -0.046 &
     -0.038 &
            \tabularnewline
\hline\hline
\end{tabular}  
\end{table}
\endgroup

\begingroup
\squeezetable
\begin{table}
\caption{Deviation (experiment--theory) for computed electron affinities (eV)\label{tab:EA}}

\begin{tabular}{lrrr}
\hline\hline
  molecule &
         W2 &
         W3 &
 Expt. uncertainty \tabularnewline
\hline
      B &
       0.015 &
       0.005 &
       0.00003 \tabularnewline
      C &
       0.007 &
      -0.007 &
       0.0003 \tabularnewline
      O &
       0.012 &
      -0.003 &
       0.000003 \tabularnewline
      F &
      -0.002 &
      -0.006 &
       0.000004 \tabularnewline
      Al &
       0.020 &
       0.004 &
       0.00005 \tabularnewline
      Si &
       0.010 &
      -0.001 &
       0.000006 \tabularnewline
      P &
       0.015 &
       0.005 &
       0.0003 \tabularnewline
      S &
       0.008 &
       0.003 &
       0.000001 \tabularnewline
      Cl &
       0.002 &
       0.004 &
       0.00006 \tabularnewline
      C$_2$ &
      0.031 &
      0.001 &
      0.008 \tabularnewline
        CH &
      0.029 &
      0.019 &
      0.008 \tabularnewline
       CH$_2$ &
      0.002 &
     -0.001 &
      0.006 \tabularnewline
       CH$_3$ &
      0.034 &
      0.029 &
      0.030 \tabularnewline
       Cl$_2$ &
      0.004 &
      0.004 &
      0.200 \tabularnewline
        CN &
     -0.026 &
     -0.001 &
      0.005 \tabularnewline
        NH &
      0.008 &
     -0.005 &
      0.004 \tabularnewline
       NH$_2$ &
      0.007 &
      0.006 &
      0.037 \tabularnewline
        NO &
     -0.001 &
     -0.003 &
      0.005 \tabularnewline
        O$_2$ &
     -0.003 &
     -0.004 &
      0.007 \tabularnewline
        OF &
     -0.009 &
      0.004 &
      0.006 \tabularnewline
        OH &
     -0.001 &
     -0.004 &
      0.000 \tabularnewline
        PH &
      0.010 &
      0.003 &
      0.010 \tabularnewline
       PH$_2$ &
      0.013 &
      0.009 &
      0.010 \tabularnewline
        PO &
     -0.002 &
      0.006 &
      0.010 \tabularnewline
        S$_2$ &
     -0.018 &
     -0.015 &
      0.040 \tabularnewline
        SH &
      0.008 &
      0.009 &
      0.002 \tabularnewline
      SiH$_2$ &
      0.039 &
      0.030 &
      0.022 \tabularnewline
       SiH &
      0.031 &
      0.021 &
      0.009 \tabularnewline
      SiH$_3$ &
      0.011 &
     -0.001 &
      0.014 \tabularnewline
\hline
 mean abs. &
     0.0135 &
     0.0076 &
            \tabularnewline
       RMS &
     0.0173 &
     0.0109 &
            \tabularnewline
    max(+) &
       SiH$_2$ &
       SiH$_2$ &
            \tabularnewline
           &
      0.039 &
      0.030 &
            \tabularnewline
    max(-) &
         CN &
         S$_2$ &
            \tabularnewline
           &
     -0.026 &
     -0.015 &
            \tabularnewline
\hline\hline
\end{tabular}  
\end{table}
\endgroup

\begingroup
\squeezetable
\begin{table}
\caption{Performance of W2 and W3 theory for total atomization energies. Deviations given are experiment--theory (kcal/mol)\label{tab:TAE}}
\begin{tabular}{lrrrrr}
\hline\hline
  molecule &
 Error in W2 &
 Error in W3 &
 Error in W3A &
 Error in W3K &
 Expt. uncertainty \tabularnewline
\hline
      C$_2$H$_2$ &
       0.42 &
       0.43 &
       0.43 &
       0.53 &
       0.24 \tabularnewline
      C$_2$H$_4$ &
      -0.19 &
      -0.19 & 
      -0.16 &
      -0.08 &
       0.24 \tabularnewline
       CH$_3$ &
      -0.21 &
      -0.27 &
      -0.25 &
      -0.25 &
       0.10 \tabularnewline
       CH$_4$ &
      -0.11 &
      -0.14 &
      -0.11 &
      -0.09 &
       0.14 \tabularnewline
        CH &
      -0.08 &
      -0.23 &
      -0.23 &
      -0.23 &
       0.23 \tabularnewline
       CO$_2$ &
       0.14 &
      -0.13 &
      +0.04 &
      +0.09 &
       0.12 \tabularnewline
      H$_2$CO &
      -0.27 &
      -0.41 &
      -0.31 &
      -0.26 &
       0.12 \tabularnewline
       H$_2$O &
      -0.04 &
      -0.16 &
      -0.08 &
      -0.08 &
       0.12 \tabularnewline
        HF &
       0.02 &
      -0.10 &
      -0.02 &
      -0.04 &
       0.17 \tabularnewline
       HNO &
       0.38 &
      -0.11 &
      -0.06 &
      +0.03 &
       0.06 \tabularnewline
       NH$_3$ &
      -0.03 &
      -0.12 &
      -0.09 &
      -0.08 &
       0.13 \tabularnewline
        N$_2$O &
       1.20 &
       0.51 &
       0.57 &
       0.68 &
       0.10 \tabularnewline
        NO$_2$ &
       1.16 &
       0.05 &
       0.18 &
       0.32 &
       0.10 \tabularnewline
        NO &
       0.47 &
       0.09 &
       0.15 &
       0.18 &
       0.03 \tabularnewline
        O$_2$ &
       0.64 &
       0.02 &
       0.11 &
       0.10 &
       0.04 \tabularnewline
        O$_3$ &
       3.01 &
       0.38 &
       0.36 &
       0.67 &
       0.03 \tabularnewline
        N$_2$ &
       0.36 &
       0.06 &
       0.06 &
       0.15 &
       0.04 \tabularnewline
        CO &
       0.12 &
      -0.03 &
      +0.04 &
       0.10 &
       0.12 \tabularnewline
        F$_2$ &
       0.60 &
      -0.09 &
      +0.01 &
       0.04 &
       0.10 \tabularnewline
       Cl$_2$ &
      -0.21 &
      -0.14 &
      -0.15 &
       0.04 &
       0.00 \tabularnewline
       ClF &
       0.10 &
      -0.10 &
      -0.01 &
       0.05 &
       0.01 \tabularnewline
        CS &
       0.26 &
       0.21 &
       0.12 &
       0.32 &
       0.23 \tabularnewline
       H$_2$S &
      -0.39 &
      -0.43 &
      -0.42 &
      -0.36 &
       0.12 \tabularnewline
       HCl &
      -0.05 &
      -0.06 &
      -0.06 &
       0.00 &
       0.02 \tabularnewline
      HOCl &
      -0.16 &
      -0.30 &
      -0.23 &
      -0.14 &
       0.12 \tabularnewline
       PH$_3$ &
       0.01 &
      -0.07 &
      -0.25 &
      -0.04 &
       0.41 \tabularnewline
        SO &
       0.01 &
      -0.14 &
      -0.02 &
      -0.04 &
       0.04 \tabularnewline
       SO$_2$ &
      -0.28 &
      -0.78 &
       ---   &
      -0.46 &
       0.08 \tabularnewline
       OCS &
      -0.21 &
      -0.41 &
       ---   &
      -0.21 &
       0.48 \tabularnewline
      ClCN &
       0.38 &
       0.31 &
       ---   &
       0.50 &
       0.48 \tabularnewline
\hline
  mean signed error$^a$ &
      0.24(0.26) & -0.08(-0.04) & (-0.01) & +0.05(0.07) & \\
  mean abs. error &
      0.40(0.36) &
      0.22(0.16) & (0.16) & 0.20(0.18) & 0.15$b$ \tabularnewline
       RMS error &
      0.70(0.72) &
      0.28(0.23) & (0.22) & 0.28(0.26) \tabularnewline
   largest pos. dev. &
         O$_3$ &
       N$_2$O & N$_2$O & N$_2$O \tabularnewline
            &
       3.01 &
       0.51 & 0.57  & 0.68 \tabularnewline
   largest neg. dev. &
        H$_2$S &
        SO$_2$ & H$_2$S & SO$_2$ \tabularnewline
           &
      -0.39 &
      -0.78 & -0.42 & -0.46 \tabularnewline
\hline\hline
\end{tabular}  

(a) Error statistics in parentheses are exclusive of SO$_2$, OCS, and ClCN. 

(b) average experimental uncertainty

\end{table}
\endgroup

\begingroup
\squeezetable
\begin{table}
\caption{Comparison of W3, W4a, and W4b for TAEs (kcal/mol)\label{tab:W4}}
\begin{tabular}{l|rrrrr}
\hline\hline
  molecule &
         W2 &
         W3 &
 W4a &
   W4b &
 uncertainty \tabularnewline
\hline
      C$_2$H$_2$ &
       0.42 &
       0.43 &
       0.29 &
       0.32 &
       0.24 \tabularnewline
       CH$_3$ &
      -0.21 &
      -0.27 &
      -0.16 &
      -0.15 &
       0.10 \tabularnewline
       CH$_4$ &
      -0.11 &
      -0.14 &
      -0.09 &
      -0.08 &
       0.14 \tabularnewline
        CH &
      -0.08 &
      -0.23 &
      -0.21 &
      -0.21 &
       0.23 \tabularnewline
       H$_2$O &
      -0.04 &
      -0.16 &
       0.09 &
       0.15 &
       0.12 \tabularnewline
        HF &
       0.02 &
      -0.10 &
       0.09 &
       0.16 &
       0.17 \tabularnewline
       NH$_3$ &
      -0.03 &
      -0.12 &
       0.11 &
       0.11 &
       0.13 \tabularnewline
        NO &
       0.47 &
       0.09 &
       0.08 &
       0.18 &
       0.03 \tabularnewline
        O$_2$ &
       0.64 &
       0.02 &
       0.09 &
       0.26 &
       0.04 \tabularnewline
        N$_2$ &
       0.36 &
       0.06 &
       0.16 &
       0.26 &
       0.04 \tabularnewline
        CO &
       0.12 &
      -0.03 &
      -0.17 &
      -0.10 &
       0.12 \tabularnewline
        F$_2$ &
       0.60 &
      -0.09 &
      -0.02 &
       0.13 &
       0.10 \tabularnewline
       Cl$_2$ &
      -0.21 &
      -0.14 &
       0.05 &
       0.06 &
       0.00 \tabularnewline
       ClF &
       0.10 &
      -0.10 &
      -0.14 &
      -0.06 &
       0.01 \tabularnewline
        CS &
       0.26 &
       0.21 &
      -0.20 &
      -0.23 &
       0.23 \tabularnewline
       H$_2$S &
      -0.39 &
      -0.43 &
      -0.43 &
      -0.42 &
       0.12 \tabularnewline
       HCl &
      -0.05 &
      -0.06 &
       0.01 &
       0.02 &
       0.02 \tabularnewline
       PH$_3$ &
       0.01 &
      -0.07 &
       0.17 &
       0.16 &
       0.41 \tabularnewline
        SO &
       0.01 &
      -0.14 &
      -0.15 &
      -0.05 &
       0.04 \tabularnewline
\hline
  mean abs &
      0.224 &
      0.154 &
      0.142 &
      0.170 &
            \tabularnewline
       RMS &
      0.302 &
      0.194 &
      0.172 &
      0.197 &
            \tabularnewline
   max (+) &
       C$_2$H$_2$ &
       C$_2$H$_2$ &
       C$_2$H$_2$ &
       C$_2$H$_2$ &
            \tabularnewline
           &
       0.42 &
       0.43 &
       0.29 &
       0.32 &
            \tabularnewline
    max(-) &
        H$_2$S &
        H$_2$S &
        H$_2$S &
        H$_2$S &
            \tabularnewline
           &
      -0.39 &
      -0.43 &
      -0.43 &
      -0.42 &
            \tabularnewline
\hline\hline
\end{tabular}

\end{table}
\endgroup


\begin{thebibliography}{20}
\bibitem{closedMOLPRO} The easiest way to break down the CCSD correlation energyin singlet and triplet
pair components using an unmodified copy of MOLPRO, in a closed-shell case, is to forcibly run the
CCSD part as open-shell since the closed-shell CCSD code does not generate pair correlation energies. 
CPU time requirements will approximately double for the affected parts of the calculation.
The program will report $E_{\alpha\alpha}$, $E_{\beta\beta}$, and $E_{\alpha\beta}$, from which 
$E^S=E_{\alpha\beta}-E_{\beta\beta}$ and $E^T=E_{\alpha\alpha}+2E_{\beta\beta}$ reproduces Klopper's partitioning\cite{KlopperExtrap}.
\end{thebibliography}
\end{document}